\documentclass[10pt,a4paper]{article}

\usepackage[left=2cm,right=2.5cm, top=2cm,bottom=3cm]{geometry}
\usepackage[utf8]{inputenc}
\usepackage[english,russian]{babel}
\usepackage[pdftex]{graphicx}
\usepackage{amsmath}
\usepackage{amsthm}
\usepackage{amssymb}
\usepackage{amsfonts}
\usepackage{eucal}
\usepackage{setspace}
\usepackage{mathtools}
\usepackage{afterpage}
\usepackage[svgnames]{xcolor}
\usepackage{cancel}
\usepackage[breaklinks,colorlinks,allcolors=DarkCyan,unicode,pdftitle={On the adiabatic invariance of the action of a trapped wave}]{hyperref}

\usepackage{authblk}
\usepackage{times}
\usepackage[sort&compress]{natbib}

\theoremstyle{remark}
\newtheorem{remark}{Remark}

\providecommand{\keywords}[1]
{{
  \small	
  \textbf{\itshape{Keywords ---}} #1
}}

\renewcommand{\=}{\stackrel{\mbox{\scriptsize def}}{=}}
\newcommand{\I}{{\mathrm{i}}}
\renewcommand{\d}{\mathrm{d}}
\renewcommand{\v}{v}
\renewcommand{\O}{\Omega}

\newcommand{\TF}{\mathcal{T}}
\newcommand{\WF}{\mathcal{W}}
\newcommand{\UF}{\mathcal{U}}

\renewcommand{\l}{\ell}

\newcommand{\Exp}[1]{\operatorname{exp}\left(#1\right)}
\newcommand\EXP[1]{\mathrm e^{#1}}
\newcommand{\const}{\text{const}}

\newcommand{\F}[1]{{#1}_F}

\newcommand{\abs}[1]{\left| #1 \right|}

\def\cc{\mathrm{c.c.}}

\newcommand\pd[2]{\frac{\partial#1}{\partial#2}}

\newcommand\dd[2]{\frac{\d #1}{\d#2}}

\newcommand\ttp{\hat t}

\newcommand{\PARTIAL}{(\cdot)}

\newcommand{\mathcalC}{C}
\newcommand{\et}{{\epsilon t}}
\newcommand\pbaru{{\{\bar u\}}}
\newcommand\baru{{\bar u}}

\numberwithin{equation}{section}
\numberwithin{remark}{section}

\begin{document}

\selectlanguage{english}

%
%

\title{On the adiabatic invariance of the action of a trapped wave}

\author[1]{Ekaterina V. Shishkina}
\author[1]{Serge N. Gavrilov}
\affil[1]{Institute for Problems in Mechanical Engineering RAS, St.~Petersburg, Russia}

\maketitle

\selectlanguage{english}

\begin{abstract} 
  Recently, it has been shown
  (Gavrilov et al., Nonlinear Dyn, 112, 2024)
  that
  in a linear solid  discrete-continuous system 
  with several slowly time-varying parameters, 
  the amplitude 
  of a strongly localized mode (a trapped wave)
  can be calculated as a function of current parameter values and does not
  depend on the history of the parameter change. This result allows us to introduce
  the adiabatic invariant for such a system according to the general definition as a quantity that remains approximately constant 
  if the parameters vary slowly.
  In this paper, we show that,
  defined in this manner, the adiabatic invariant can be calculated as the
  ratio of the total energy of the trapped wave to its frequency. This yields a significantly
  simplified approach to solving a class of problems concerning localized oscillation
  of continuous systems with discrete inclusions, although 
  the definition of the wave energy can be ambiguous. 
  Thus, we can consider the newly
  introduced adiabatic invariant as a straightforward generalization of the concept known to
  Hamiltonian systems. Finally, we introduce an effective Hamiltonian system,
  which is characterized by the same adiabatic invariant as the trapped wave.
  This yields another highly straightforward approach to deriving the amplitude evolution law,
  although further investigation is required.

\end{abstract}


\tableofcontents

\section{Introduction}

  An adiabatic invariant \cite{Arnold2009,Henrard1993} is a quantity that stays approximately constant in
  a system with properties depending on a slowly time-varying parameter. In classical mechanics,
  adiabatic invariants are usually discussed in the framework of Hamiltonian formalism for systems with
  time-varying Hamiltonians. It is
  well-known that for a single-degree-of-freedom Hamiltonian system, the
  canonical action variable of the action--angle pair is an adiabatic invariant. In particular, for a linear oscillator, the
  action equals the ratio of the energy to the frequency. 

  The theory of the wave action, which was developed by Whitham
  et al., is also known in the literature \cite{Whitham1999,Buehler2009,Bretherton1966,Whitham1965,Whitham1967}; see also papers in the  special issue 
  \cite{Lighthill1967full}, and discussions in \cite{Babich2002}.
  This theory was developed keeping in mind the
  water waves and is applicable for a special kind of solutions referred to as
  a wave-train (a slightly modulated running plane wave with infinite energy).
  In the framework of
  this theory, it is shown that the wave action, i.e., the ratio of the energy density to the frequency, is also, in a
  certain sense, the adiabatic invariant.
  Namely, it is possible to obtain the conservation law for the invariant in the form 
  of a partial differential equation (PDE) of the first order.  
  These results may allow one to
  calculate the amplitude of the wave-train propagating along a medium. 

  To the best of our knowledge, there are still no studies concerning
  adiabatic invariants related to linear
  localized waves, also known as trapped modes. Trapped modes initially were discovered by Ursell
  \cite{ursell1951trapping} and can exist in various fluid and solid media
  and constructions \cite{Ind-book-R2E,Mishuris2020,mciver2003excitation,Indeitsev2006,indeitsev2012motion,kaplunov1995simple,gavrilov2002etm,Glushkov2011a}. Recently, it has been shown \cite{Gavrilov2024nody}
  that
  in a linear discrete-continuous system with
  time-varying parameters, 
  the amplitude 
  of a strongly
  \cite{luongo2001mode,Luongo1992}
  localized mode that is a linear wave with finite energy
  can be calculated as a function of current parameter values and does not
  depend on the history of the parameter change. This allows one to introduce
  the adiabatic invariant for such a system according to the general definition as a quantity 
  that remains approximately constant in
  a system with slowly time-varying parameters. In this paper, we show that,
  defined in this manner, the adiabatic invariant can be calculated as the
  ratio of localized mode energy to its frequency. This result yields a significantly
  simplified approach to solve a class of problems concerning localized oscillation
  of continuous systems with discrete inclusions, although 
  the definition of ``mode energy'' can be ambiguous. 
  We can consider a newly
  introduced adiabatic invariant as a generalization of the concept known to
  Hamiltonian systems, perhaps a more straightforward one than the action of a wave-train.
 
In the paper, we consider a solid 
discrete-continuous mechanical system, where a strongly localized oscillation
mode is possible. The system is a taut string on the Winkler foundation
coupled to a discrete mass--spring sub-system.
The existence of a trapped mode for the simplest
particular case of such a system was demonstrated in
\cite{Abramyan1994,kaplunov1995simple,Glushkov2011a}. Due to the existence
of the trapped mode, a localized non-vanishing oscillation 
\cite{kaplunov1986torsional}
is the
leading-order term of the solution for the corresponding non-stationary
problem. We refer to this phenomenon as the localization of non-stationary waves \cite{Shishkina2023jsv}. The
corresponding non-stationary solution for such a system with multiple slowly
time-varying parameters is obtained in our previous paper \cite{Gavrilov2024nody} by
an asymptotic approach which is quite similar to the space-time ray method \cite{Babich2009,Babich2002,Babich2025,Hinch2002} and is inspired \cite{gavrilov2002etm}
by Nayheh's method of multiple scales \cite{nayfeh2008perturbation}. 
The initial conditions are satisfied by matching this solution with that
obtained via the method of stationary phase applied to the corresponding
zeroth-order equations.

The paper is organized as follows. In Sect.~\ref{sect-fixed} and
\ref{sect-moving}, we consider two particular cases of the problem discussed
above; namely, the problems with a mass-spring system at fixed and variable positions,
respectively. The solutions for these cases are obtained in \cite{Gavrilov2024nody}
within the framework of the asymptotic approach common to both problems. However, in
the current paper, we consider these two cases separately. The problem in Sect.~\ref{sect-fixed} 
is treated as a reference problem. In Sects.~\ref{sect-form}, we provide the 
mathematical formulation of the problem. In
Sects.~\ref{sect-trapped-fixed}--\ref{sect-asy-fixed}, we provide necessary
preliminaries to deal with the asymptotic solution obtained in \cite{Gavrilov2024nody}.
Knowing the asymptotic solution, 
in Sect.~\ref{sect-adibatic-fixed}, we introduce the corresponding adiabatic invariant and
prove that the invariant is a function of the action of the trapped wave, i.e., the ratio of the
mode energy to its frequency. 
In Sect~\ref{sect-moving}, we apply this approach to an extended problem in which the discrete sub-system can move along 
the continuous one.
Again, in Sect.~\ref{sect-form-moving}, we provide the 
mathematical formulation of the problem. In
Sects.~\ref{sect-trapped-moving}--\ref{sect-asy-moving}, we provide necessary
preliminaries to deal with the asymptotic solution obtained in \cite{Gavrilov2024nody}. 
In Sect.~\ref{sect-adibatic-moving},
we attempt to re-derive this solution by postulating that the action of the trapped wave is an adiabatic invariant, and we demonstrate the difficulties associated with this approach.
These difficulties arise from the ambiguity in selecting the appropriate energy functional for calculating the action; see also
Appendix~\ref{App-BAD}.
Finally, in
Sect.~\ref{sect-discuss}, 
we introduce an effective Hamiltonian system which possesses the same
adiabatic invariant, namely, the action, as the original system.
The original system inherits some properties from this Hamiltonian system. This allows 
us to solve the non-stationary perturbed problem without any asymptotic calculations,
given the solution of the non-stationary unperturbed problem with time-independent parameters.
The motions in the effective Hamiltonian system are investigated in detail in Appendix~\ref{App-1d}.

\section{Discrete mass-spring inclusion at a fixed position}
\label{sect-fixed}
\subsection{Mathematical formulation}
\label{sect-form}

The transverse oscillation of a taut string on the Winkler foundation with a
discrete oscillatory inclusion can be described by the following coupled
system of differential equations \cite{Gavrilov2024nody}:
\begin{gather}
\label{eq2}
\frac{\d }{\d t}
\left( M \frac{\d \UF}{\d t} \right)
+ K \UF = -P(t) +p(t),
\\
\label{eq1}
\TF \, \frac{\partial^2 u}{\partial^2 x}
- \frac{\partial }{\partial t}
\left(
\rho \frac{\partial u}{\partial t} \right)
- k u 
= -P(t)\delta\big(x)
,
\end{gather}
formulated for $x\in \mathbb R$, $t\in(0,+\infty)$. 
Here, $t$ is time, 
$x$ is the spatial co-ordinate,
$u(x,t)$ is the displacements, 
\begin{equation}
\mathcal U=u(0,t)
\end{equation}
is the point mass displacement,
$M$ is the mass of the discrete inclusion, $K$ is the corresponding
stiffness,
$k$ is the elastic foundation stiffness, $\rho$ is
the mass density, $\mathcal T$ is the string tension,
$p(t)$  is a given external force,
$P(t)$ is an unknown internal interaction force between the string and the point mass.
In the paper, we assume that the external loading $p$ is a pulse of a finite
duration:
\begin{equation}
p({t})\equiv 0, \qquad {t}<0 \quad\text{or}\quad {t}>{t}_0>0
\label{pulse}
\end{equation}
for certain $t_0>0$. 
All parameters are assumed to be smooth functions of the slow time
$\epsilon t$:
\begin{equation}
\begin{gathered}	
K=K(\epsilon t),
\qquad
M=M(\epsilon t),
\\
\mathcal T=\mathcal T(\epsilon t),
\quad
\rho=\rho(\epsilon t),
\quad
k=k(\epsilon t),
\end{gathered}
\label{all-slow}
\end{equation}
such that the following restrictions are satisfied:
\begin{gather}
M\geq0,
\qquad
\mathcal T>0,
\qquad
\rho \geq0;
\qquad
k>0;
\label{restr-par-21}
\\
M>0\quad\text{or}\quad\rho>0.
\label{restr-par-22}
\end{gather}
\begin{remark} 
Restriction 
\eqref{restr-par-22} was not formulated explicitly in 
\cite{Gavrilov2024nody} but it was assumed there.
\end{remark} 
The quantity $\epsilon>0$ is the
dimensionless formal small parameter:
\begin{equation}
\epsilon=o(1).
\label{epsilon-def}
\end{equation}
The initial conditions are zero and can be formulated in the following form \cite{Vladimirov1971}:
\begin{equation}
 u\big|_{t<0}\equiv0.
 \label{ic}
\end{equation}
The boundary conditions at infinity can be
formulated as follows:
\begin{equation}
u\equiv0 \quad\text{for}\quad |x|>x_0.
\label{boundary-c-non}
\end{equation}
\subsection{Stationary unperturbed problem}
\label{sect-trapped-fixed}
Firstly, consider the corresponding stationary ($p=0$) unperturbed ($\epsilon=0$) problem. Quantities 
\eqref{all-slow} are assumed to be constants:
\begin{equation}
\begin{gathered}	
K=\const,
\qquad
M=\const,
\\
\mathcal T=\const,
\quad
\rho=\const,
\quad
k=\const.
\end{gathered}
\label{all-constants}
\end{equation}

Provided that certain conditions, which we call the localization conditions,  are satisfied,
 a unique linear strongly localized standing wave with finite energy (referred to as  
 a trapped mode) exists in the system 
 \cite{Gavrilov2024nody}:
\begin{equation}
u=\bar u\=|C|\Exp{-S\big(\Omega_0\big)|x|}\cos(\Omega_0 t-\arg C).
\label{eq:solution_inhomogeneous_stationary_problem-ext}
\end{equation}
Here, $C$ is an arbitrary complex constant,
\begin{equation}	
\label{S0}
S(\O)=
{\sqrt\frac{k-\rho\O^2}{\TF}}.
\end{equation}
The exact form of the localization conditions
is not necessary for this paper. 
In Eq.~\eqref{eq:solution_inhomogeneous_stationary_problem-ext},
$\Omega_0>0$ is the frequency of the localized oscillation, which satisfies 
the frequency equation
\begin{equation}
2\sqrt{\TF (k-\rho\O_0^2)}={M\O_0^2-K}.
\label{fr_eq-1}
\end{equation}

\subsection{Asymptotic solution of the non-stationary unperturbed problem by the method of
stationary phase}
\label{sect-st-ph}
Next, consider the non-stationary problem with $p\neq0$ in the zeroth-order approximation ($\epsilon=0$), where quantities 
\eqref{all-slow} are assumed to be constants:
\begin{equation}
\begin{gathered}	
K=K(0),
\qquad
M=M(0),
\\
\mathcal T=\mathcal T(0),
\quad
\rho=\rho(0),
\quad
k=k(0),
\end{gathered}
\label{all-slow0}
\end{equation}
equal the corresponding initial values at $t=0$. We assume that the
localization conditions are satisfied for initial values 
\eqref{all-slow0}
of the system parameters.

For a system with non-zero $p(t)$ satisfying 
Eq.~\eqref{pulse}, the leading-order term of the asymptotic solution for large times 
\begin{equation}
t\gg t_0
\label{tt0}
\end{equation}
can be obtained by
the method of stationary phase. According to \cite{Gavrilov2024nody},
the leading-order term has the following form:
\begin{equation}
u=|C(0)|\Exp{-S\big(\Omega_0(0)\big)|x|}\cos(\Omega_0(0) t-\arg C(0)).
\label{eq:solution_inhomogeneous_stationary_problem-ext-zero}
\end{equation}
Here,
$\Omega_0(0)>0$ is the trapped mode frequency, which satisfies 
the frequency equation
\eqref{fr_eq-1},
where all involved system parameters are taken at $t=0$. The absolute value
and the argument of the complex constant $C(0)$ in 
Eq.~\eqref{eq:solution_inhomogeneous_stationary_problem-ext-zero} are
\begin{gather}
|C(0)|=\big|C_0(0)\F{p}\big(\O_0(0)\big)\big|,
\label{CCCC}
\\
|C_0(0)|=
\left.\frac{
    \sqrt{k-\rho\O_0^2}
  }{\O_0\left(\sqrt\TF\rho+M \sqrt{k-\rho\O_0^2}\right)}\right|_{t=0},
\label{amp0}
\\
\arg C(0)=\arg \F{p}\big({\O_{0}(0)}\big)
+
\arg C_0
,
\label{argC}
\\
\arg C_0(0)=
\frac\pi2
,
\label{Co-value}
\end{gather}
i.e.,
\begin{gather}
\UF(t) =
|C(0)|\cos\big(\I \phi - \arg C(0)\big)+\text{o}(1), \quad t\gg t_0;
\label{eq:solution_inhomogeneous_stationary_problem}
\\
\phi=-\I \Omega_0(0) t.
\label{phiOt}
\end{gather}
Here, the symbol $p_F\big(\O_0(0)\big)$ denotes the Fourier transform of $p(t)$:
\begin{equation}
p_F=\int_{-\infty}^{+\infty}p\EXP{\I\Omega t}\,\d t
\end{equation}
calculated
at $\Omega=\Omega_0(0)$. Thus, $C(0)=C_0(0)$ for $p(t)=\delta(t)$.

\subsection{Asymptotic solution of the non-stationary perturbed problem by the space-time ray
method}
\label{sect-asy-fixed}
We now allow the system parameters to vary according to Eq.~\eqref{all-slow}, ensuring that the localization conditions remain satisfied for all
$\epsilon t$.
In  \cite{Gavrilov2024nody} it is shown that the solution has the form
typical of the WKB approach:
\begin{gather}
\UF(t) =
|\mathcal C(\epsilon t)|
\cos\big(\I\phi - \arg \mathcal C\big)+\text{o}(1), 
\label{sol-eg0}
\\
|\mathcal C|=|\mathcal C(\epsilon t)|=|C(0)|{\frac{|\mathcal C_0(\epsilon t)|}{|\mathcal C_0(0)|}},
\label{amp}
\\
\arg \mathcal C=\arg C(0),\qquad
\arg \mathcal C_0=\arg C_0(0),
\\
\phi=
{-}\I\int_{0}^{t}
 \Omega_0\big(\epsilon\hat t\big)\,\d \hat t,
\label{phi-def}
\\
|\mathcal C_0|=
|\mathcal C_0(\epsilon t)|\=
\sqrt{
\frac{
    \sqrt{k-\rho\O_0^2}
}{\O_0\left(\sqrt\TF\rho+M \sqrt{k-\rho\O_0^2}\right)}
}.
\label{amp-expl}
\end{gather}
Here, 
\begin{equation}
t/t_0=O(\epsilon^{-1}).
\label{tt0>}
\end{equation}
Again, $\mathcal C(\epsilon t)=\mathcal C_0(\epsilon t)$ for $p(t)=\delta(t)$.
The structure of the right-hand side of Eq.~\eqref{amp} guarantees that 
\begin{itemize} 
  \item
  As $\epsilon t\to0$,
  Eq.~\eqref{sol-eg0} reduces to
  Eq.~\eqref{eq:solution_inhomogeneous_stationary_problem} due to the
  matching procedure used in \cite{Gavrilov2024nody};\footnote{Thus, large times for
  asymptotic procedure from 
  Sect.~\ref{sect-st-ph} are small times for the asymptotics considered now.}
  \item
    The amplitude $|\mathcal C|$ for $t/t_0=O(\epsilon^{-1})$ is proportional to
    the right-hand side of Eq.~\eqref{amp-expl}:
\begin{equation}
  |\mathcal C(\et)|\propto|\mathcal C_0(\et)|
    .
  \label{propto}
\end{equation}
\end{itemize} 

\begin{remark} 
  The asymptotic ansatz that was used in \cite{Gavrilov2024nody} to obtain these
  results has a structure that is significantly more complicated than the one presented in
  Eqs.~\eqref{sol-eg0}--\eqref{phi-def} and takes into account the alteration of
  the solution for $x\neq0$ in the case $\epsilon>0$. 
\end{remark} 

\begin{remark} 
Formula \eqref{amp0}, on the one hand, and Eqs.~\eqref{amp}, \eqref{propto}, on the other hand,  had been absolutely independently
 obtained in \cite{Gavrilov2024nody} by two entirely different approaches. However, 
the resulting connection 
  \begin{equation}
    |\mathcal C_0(\epsilon t)|=\sqrt{|C_0(\epsilon t)|}
    \label{C0C0}
  \end{equation}
has not been previously identified.
  Here, the quantity $|C_0(\epsilon t)|$ is
  calculated by the same formula as in Eq.~\eqref{amp0}, except that the system parameters and the 
  frequency $\Omega_0$ 
  are now functions of $\epsilon t$.
  The identity 
  \eqref{C0C0} is discussed in what follows; see Sect.~\ref{sect-discuss}.
\end{remark}

\subsection{Adiabatic invariance of the action of the trapped wave}
\label{sect-adibatic-fixed}
According to Eqs.~\eqref{amp}, \eqref{amp-expl},
\begin{equation}
 I=
 \frac{|\mathcal C|}{{|\mathcal C_0|}}
 =
|\mathcal C|
\sqrt{
 \frac
{\O_0\left(\sqrt\TF\rho+M \sqrt{k-\rho\O_0^2}\right)}
{
    \sqrt{k-\rho\O_0^2}
}}
 =\frac{|C(0)|}{|\mathcal C_0(0)|}
=
\const,
\label{I-intro}
\end{equation} 
for times
\eqref{tt0>}.
Thus, according to the direct asymptotic analysis presented in \cite{Gavrilov2024nody},
we have found that $I$ or any function $f(I)$ of the single variable $I$
such that $f'\neq0$ 
is the adiabatic invariant. 
Here, we introduce this concept for the system under consideration as a quantity that remains approximately constant in systems with slowly time-varying parameters, consistent with the general definition
as a quantity that remains approximately constant in
a system with several slowly time-varying parameters. 

Let us show that the action of the trapped wave
\begin{equation}
  J\=\left.\frac {E\{\bar u\}}{\Omega_0}
    \right|_{C=\mathcal C}
\end{equation}
is a function of the adiabatic
invariant $I$:
\begin{equation}
  J
    =f(I)
\end{equation}
for a certain function $f$. Here, $E$ denotes the total energy functional:
\begin{equation}
E\pbaru=E_{\mathrm c}\pbaru+E_{\mathrm d}\pbaru;
	\label{energy-unpert}
\end{equation}
where
\begin{gather}
  E_{\mathrm c}\pbaru=\frac12\int_{-\infty}^{+\infty} 
  \bigg(
    \rho\left(\pd{\baru}{t}\right)^2 + \TF\left(\pd{\baru}{x}\right)^2  + {k \baru^2} 
\bigg )\, \d x,
	\label{energy-unpert-c}
  \\
  E_{\mathrm d}\pbaru=\frac12\bigg(M \left(\frac {\d \baru}{\d t}\right)^2 + 	K \baru^2\bigg) \bigg\vert_{x=0}
	\label{energy-unpert-d}
\end{gather}
are the total energies for the continuous and discrete sub-systems,
respectively, and $\baru$ is defined by Eq.~\eqref{eq:solution_inhomogeneous_stationary_problem-ext}.
Substituting Eq.~\eqref{eq:solution_inhomogeneous_stationary_problem-ext}
into Eqs.~\eqref{energy-unpert-c}, \eqref{energy-unpert-d}, respectively, one gets
\begin{gather}
  \frac{2 E_{\mathrm c}}{\mathcalC^2}
  =\frac{\rho\Omega_0^2(1-\cos 2\psi)}{2S(\Omega_0)}
  +
  \frac{(\mathcal T-\rho v^2)S(\Omega_0)(1+\cos2\psi)}{2}
  +
\frac {k(1+\cos 2\psi)}{2S(\Omega_0)},
\label{Ec}
\\
  \frac{2E_{\mathrm d}}{\mathcalC^2}=M\Omega_0^2\sin^2\psi+ K\cos^2\psi,
\label{Ed}
\end{gather}
where 
\begin{equation}
 \psi
 =\Omega_0t-\arg C.
\end{equation}
Substituting Eqs.~\eqref{S0}, \eqref{Ec}, \eqref{Ed} into Eq.~\eqref{energy-unpert} results in:
\begin{gather} 
  \frac{2 E_{\mathrm c}}{\mathcalC^2}
  +
  \frac{2 E_{\mathrm d}}{\mathcalC^2}
  =
  \left(
    \frac{\Omega_0^2(\sqrt{\TF}\rho+M\sqrt{k-\rho\Omega_0^2})}{\sqrt{k-\rho\Omega_0^2}}
  \right.
  +
  \left(K-M\Omega_0^2+
2\sqrt{\TF (k-\rho\O_0^2)}\right)
\cos2\psi
  \Bigg).
\end{gather} 
The second term in the right-hand side of the last equation is zero according
to frequency equation 
\eqref{fr_eq-1}. 
%
Thus,
\begin{equation}
  J=
  \left.
  \frac {E\pbaru}{\Omega_0}
    \right|_{C=\mathcal C}
  =\frac{\mathcal C^2}{2}\left( \frac{\Omega_0(\sqrt{\TF}\rho+M\sqrt{k-\rho\Omega_0^2})}{\sqrt{k-\rho\Omega_0^2}} \right)
  =\frac{I^2}2=
	f(I)
	\label{inv-1}
\end{equation}
due to Eq.~\eqref{I-intro}, i.e., the trapped mode action $J$
is really an adiabatic invariant.

Conversely, we can show now that Eqs.~\eqref{amp-expl}, \eqref{propto} can be obtained by an alternative method significantly
simpler than the  asymptotic
approach used in \cite{Gavrilov2024nody}, which is based on the space-time ray method. Indeed,  postulating that 
the action  $J$ of the trapped wave
is an adiabatic
invariant clearly leads to Eqs.~\eqref{amp-expl}, \eqref{propto}:
\begin{equation}
 J=\const
 \quad\Longrightarrow\quad
  |\mathcal C|\propto
  \sqrt{
\frac{
    \sqrt{k-\rho\O_0^2}
}{\O_0\left(\sqrt\TF\rho+M \sqrt{k-\rho\O_0^2}\right)}
  }
\end{equation}
due to Eq.~\eqref{inv-1}.

\begin{remark} 
  \label{remark-ham2}
We can intuitively think that  an effective single-degree-of-freedom Hamiltonian system
can be corresponded somehow 
to the solution of the non-stationary problem with
time-varying parameters formulated in Sect.~\ref{sect-form} for times \eqref{tt0>},
see Sect.~\ref{sect-discuss} for further discussion.

\end{remark} 


\section{Moving discrete mass-spring inclusion}
\label{sect-moving}
\subsection{Mathematical formulation}
\label{sect-form-moving}
Following  
\cite{Gavrilov2024nody},
consider now a kind of moving load problem
\cite{fryba1972vibration}, namely, an oscillatory inclusion moving along  the string on the Winkler 
foundation. The equations of motion are Eq.~\eqref{eq2} and the following
one: 
\begin{gather}
\label{eq1-l}
\TF \, \frac{\partial^2 u}{\partial^2 x}
- \frac{\partial }{\partial t}
\left(
\rho \frac{\partial u}{\partial t} \right)
- k u 
= -P(t)\delta\big(x-\l(t)\big)
.
\end{gather}
The point mass displacement
$\mathcal U$ in Eq.~\eqref{eq2} now is
\begin{equation}
\mathcal U=u(\ell(t),t).
\end{equation}
Here, $\ell(t)$ is the discrete oscillator position along the string:
\begin{equation}
\label{l(t)}
{\l}(t) = {\l}(0) + \int_0^t {\v}(\epsilon\ttp)\,\d \ttp,
\end{equation}
$\l(0)$ is a given initial position,
\begin{equation}
v=v(\epsilon t)
\label{v(et)}
\end{equation}
is the given speed of the oscillator (a given smooth, slowly time-varying function). 
Without loss of generality, we can put
\begin{equation}
\ell(0)=0.
\end{equation}
We assume  a sub-critical regime for the motion of the discrete oscillator, i.e.,
for all $t>0$ the following inequality must be satisfied:
\begin{equation}
\label{v(t)and_c}
\abs{\v(\epsilon t)}<c(\epsilon t),
\end{equation}
where
\begin{equation}
c=\sqrt{\dfrac{\TF}{\rho}}
\label{c-def}
\end{equation}
is the local instantaneous value for the speed of the wave propagation (the speed of
sound). 

Formulae 
\eqref{pulse}--\eqref{epsilon-def}, as well as zero initial conditions
\eqref{ic} and
boundary conditions 
\eqref{boundary-c-non} remain to be applicable to the new problem.

\begin{remark} 
For $v\equiv0$, the new problem transforms into the one formulated in
Sect.~\ref{sect-form}.
\end{remark} 
\subsection{Stationary unperturbed problem}
\label{sect-trapped-moving}
Again, provided that localization conditions are satisfied,
 a (linear) localized wave (a trapped mode) exists in the system with $p\equiv0$ \cite{Gavrilov2024nody}:
\begin{equation}
u=\bar u\=|C|\Exp{-S(\Omega_0)|x-vt|}\cos\big(\Omega_0 t-B(\Omega_0)(x-vt)-\arg C\big),
\label{eq:s-ext-l}
\end{equation}
where $C$ is an arbitrary complex constant,
\begin{gather}	
\label{S}
S(\O)=
\frac{\sqrt{k\TF-k\rho\v^2-\TF\rho\O^2}}{\TF-\rho\v^2},
\\
\label{B}
B(\O)=\frac{\v\O\rho}{\TF- \rho\v^2}.
\end{gather}
The trapped mode frequency $\Omega_0>0$ satisfies the frequency equation:
\begin{equation}
2\sqrt{k\TF-k\rho\v^2-\TF\rho\O_0^2}
=M\O_0^2-K.
\label{fr_eq-1-l}
\end{equation}

\begin{remark} 
Note that for $v\neq0$, the wave described by Eq.~\eqref{eq:s-ext-l} has a finite energy, but it is not a standing wave any more. 
\end{remark} 

\subsection{Asymptotic solution of the  non-stationary unperturbed problem by the method of
stationary phase}
For a system with non-zero $p(t)$ satisfying 
Eq.~\eqref{pulse}, the leading-order term of the asymptotic solution for large times 
\eqref{tt0} can be obtained by
the method of stationary phase. The leading-order term has the following form 
\begin{equation}
u=|C(0)|\Exp{-S\big(\Omega_0(0)\big)\big|x-v(0)t\big|}\cos\Big(\Omega_0(0) t-B\big(\Omega_0(0)\big)\big(x-v(0)t\big)-\arg C(0)\Big),
\label{eq:s-ext-l0}
\end{equation}
where 
Eqs.~\eqref{CCCC},
\eqref{argC}--\eqref{phiOt} are fulfilled, and
\begin{equation}
|C_0(0)|=
\left.\frac{
    \sqrt{k\TF-k\rho\v^2-\TF\rho\O_0^2}
}{\O_0\left(\TF\rho+M \sqrt{k\mathcal T-k\rho\v^2-\TF\rho\O_0^2}\right)}
\right|_{t=0}.
\label{amp0-l}
\end{equation}

%

\subsection{Asymptotic solution of the non-stationary perturbed problem by the space-time ray
method}
\label{sect-asy-moving}
Now, let parameters of the system vary according to 
Eq.~\eqref{all-slow} such that the localization conditions are
fulfilled for any $\epsilon t$.
For times
\eqref{tt0>},
the asymptotic solution has the form \cite{Gavrilov2024nody} of Eqs.~\eqref{sol-eg0}--\eqref{phi-def}, \eqref{propto} where the amplitude 
$|C_0(0)|$
is now defined by Eq.~\eqref{amp0-l}, and 
\begin{equation}
|\mathcal C_0|= |\mathcal C_0(\epsilon t)|\=
\sqrt{
  \frac{
    \sqrt{k\TF-k\rho\v^2-\TF\rho\O_0^2}
}{\O_0\left(\TF\rho+M \sqrt{k\mathcal T-k\rho\v^2-\TF\rho\O_0^2}\right)}}.
\label{amp0-l-zzz}
\end{equation}

\subsection{Alternative solution based on the adiabatic invariance of the action of the trapped
wave}
\label{sect-adibatic-moving}

Let us try to obtain formulae 
\eqref{propto},
\eqref{amp0-l-zzz}
describing the evolution of the trapped mode
amplitude 
by the proposed
alternative way, i.e., by postulating that the action of the trapped wave is an
adiabatic invariant.

Based on physical considerations, a natural starting point is to calculate again the energy using Eq.~\eqref{energy-unpert}, where $E_{\mathrm{c}}\pbaru$ is defined by Eq.~\eqref{energy-unpert-c},
\begin{gather}
E_{\mathrm d}\pbaru=E_{\mathrm d}^{(1)}\pbaru\=
\frac12\bigg(M \left(\frac {\d \baru}{\d t}\right)^2 + 	K \baru^2\bigg) \bigg\vert_{x=\ell(t)},
	\label{energy-unpert-d1}
\end{gather}
and $\baru$ is defined by 
Eq.~\eqref{eq:s-ext-l}.
The possible alternative is to account for the kinetic energy of the
longitudinal motion of the discrete sub-system:
\begin{gather}
  E_{\mathrm d}\pbaru=E_{\mathrm d}^{(2)}\pbaru\=
  \frac12\bigg(
  \bigg(M \left(\frac {\d \baru}{\d t}\right)^2 + 	K \baru^2\bigg) \bigg\vert_{x=\ell(t)}
  +Mv^2\bigg).
	\label{energy-unpert-d2}
\end{gather}
In both cases, for $v\to0$ the energy $E_{\mathrm d}$ transforms to the one
defined by Eq.~\eqref{energy-unpert-d}. However, unfortunately, energy 
\begin{equation}
  E^{(j)}\pbaru=E_{\mathrm c}\pbaru+E^{(j)}_{\mathrm d}\pbaru,\qquad j=1,2,
	\label{energy-unpert-new}
\end{equation}
defined by Eq.~\eqref{energy-unpert-new}, \eqref{energy-unpert-c} and any of Eq.~\eqref{energy-unpert-d1} or Eq.~\eqref{energy-unpert-d2}
is not conserved even for $\epsilon=0$ provided that $v>0$.
For $\epsilon=0$, the conserved quantity
is
\begin{equation}
  \tilde E^{(j)}\pbaru={E}^{(j)}\pbaru-\int_0^t{F(\hat t)}v\,d\hat t,
	\label{energy-flux}
\end{equation}
where the second term in the right-hand side is the work of the wave pressure (or resistance) force 
\cite{rayleigh1902xxxiv,Nicolai1912-eng,nicolai1925xix,Gavrilov2016nody,
havelock1924lxx,Gavrilov(ActaMech),Slepyan2017,Slepyan2017a,Denisov2012,
Vesnitski1983,Andrianov1993,Ferretti2019},
also known as the external configurational force 
\cite{Cherepanov1985,gurtin2000cfb}:
\begin{equation}
	{F}=-\frac{\rho}{2}(c^2-v^2)
  \left[\left(\pd{\bar u}{x}\right)^2\right]
  =-\frac{1}{2}(\mathcal T-\rho v^2)
  \left[\left(\pd{\bar u}{x}\right)^2\right]
  .
	\label{conf-force}
\end{equation}
Here and in what follows, $[\mu]\equiv \mu(x+0,t)-\mu(x-0,t)$ 
for any arbitrary quantity $\mu(x,t)$. 
We can try to
calculate the corresponding actions of the trapped wave
\begin{equation}
  J^{(j)}=
  \left.
  \frac{\tilde E^{(j)}\pbaru}{\Omega_0}
  \right|_{C=\mathcal C}.
  \label{Inv-1}
\end{equation}
One can show that with this choice of the energy functional it is impossible to obtain the asymptotically correct formula 
\eqref{amp0-l-zzz}; see Appendix~\ref{App-BAD}.


To understand the structure of the required energy functional, we note that the reference asymptotic solution was derived in \cite{Gavrilov2024nody} using co-ordinates co-moving with the discrete sub-system:
\begin{equation}
\xi = x-\l(t),\qquad \tau=t. 
\label{co-moving}
\end{equation}
Using variables 
\eqref{co-moving},
Eqs.~\eqref{eq2}, \eqref{eq1-l} can be rewritten as
\begin{gather}
\label{eq2m}
\frac{\d }{\d \tau}
\left( M \frac{\d \UF}{\d \tau} \right)
+ K \UF = -P(\tau) +p(\tau),
\\
\label{eq1m}
(\TF-\rho v^2) \, \frac{\partial^2 u}{\partial^2 \xi}
+2\rho v \frac{\partial^2 u }{\partial \xi\partial \tau}
+\pd{(\rho v)}{\tau}\pd u\xi
- \frac{\partial }{\partial \tau}
\left(
\rho \frac{\partial u}{\partial \tau} \right)
- k u 
= -P(\tau)\delta(\xi)
,
\end{gather}
where
\begin{equation}
\label{U-tau}
\UF(\tau) = u(0,\tau).
\end{equation}

For the unperturbed problem, the homogeneous equation that corresponds to 
Eq.~\eqref{eq1m} is
\begin{gather}
(\TF-\rho v^2) \, \frac{\partial^2 u}{\partial^2 \xi}
+2\rho v \frac{\partial^2 u }{\partial \xi\partial \tau}
- 
\rho \frac{\partial^2 u}{\partial^2 \tau} 
- k u 
= 0
.
\label{eq1m-mov}
\end{gather}
The conserved quantity for 
Eq.~\eqref{eq1m-mov} is 
\begin{gather}
  \mathcal E_{\mathrm c}\{u\}\=
  \int_{-\infty}^{+\infty} 
  e_c
  \, \d \xi,
	\label{energy-unpert-c-mov}
  \\
  e_c\=
  \frac12\bigg (
    \rho \left(\pd{u}{\tau}\right)^2 + {(\TF-\rho v^2)}\left(\pd{u}{\xi}\right)^2 
  + {k u^2}  
  \bigg)
  =
\frac12\bigg (
  \rho\left(\pd{u}{t}\right)^2+2\rho v\pd ut \pd ux+\TF\left(\pd{u}{x}\right)^2 
  + {k u^2}
  \bigg)
  .
\end{gather}
Indeed, multiplying both sides of Eq.~\eqref{eq1m-mov}
on $\pd{u}{\tau}$ results in the following conservation law:
\begin{equation}
  \pd {e_{\mathrm c}}{\tau}=-\pd {q_{\mathrm c}}{\xi},
\end{equation}
where 
\begin{equation}
  q_{\mathrm c}=
  -(\mathcal T-\rho v^2)\bigg(\pd{u}{\xi}\bigg)\bigg(\pd{u}{\tau}\bigg)
  -\rho v \bigg(\pd u\tau\bigg)^2
\end{equation}
is the corresponding flux.
For solutions with finite energy 
\begin{equation}
 q_{\mathrm c}\to 0 
\end{equation}
as $\xi\to\infty$. Thus,
\begin{equation}
  \frac{\d \mathcal E_{\mathrm c}\{u\}}{\d \tau}=\frac{\d}{\d \tau}\int_{-\infty}^{\infty} e_c\, \d\xi=0.
\end{equation}

We define the energy functional $\mathcal E\pbaru$ evaluated
for $\bar u$ given by 
Eq.~\eqref{eq:s-ext-l} as follows:
\begin{gather}
\mathcal E\pbaru\=\mathcal E_{\mathrm c}\pbaru+\mathcal E_{\mathrm d}\pbaru,
	\label{energy-unpert-mov}
\\  
\mathcal E_{\mathrm d}\pbaru=
  \frac12\bigg(M \left(\frac {\d \baru}{\d \tau}\right)^2 + 	K \baru^2\bigg) \bigg\vert_{\xi=0}.
\end{gather}
Let us calculate the corresponding action of the trapped wave:
\begin{equation}
\mathcal J\=
\left.
\frac {\mathcal E\pbaru}{\Omega_0}
\right|_{C=\mathcal C}.
\end{equation}

One has
\begin{multline}
  \frac{2\mathcal E_{\mathrm c}}{\mathcalC^2}
  =\frac{\rho\Omega_0^2}{2S(\Omega_0)}
  \left(1-\frac{S^2(\Omega_0)\cos 2\psi}{B^2(\Omega_0)+S^2(\Omega_0)}\right)
  +
  \frac{(\mathcal T-\rho v^2)\big(B^2(\Omega_0)+S^2(\Omega_0)+S^2(\Omega_0)\cos2\psi\big)}{2S(\Omega_0)}
  \\+
  k\left(
    \frac1{2S(\Omega_0)}
    +\frac{S(\Omega_0)\cos2\psi}{2\big(B^2(\Omega_0)+S^2(\Omega_0)\big)}
    \right),
\end{multline}
\begin{gather} 
  \frac{2\mathcal E_{\mathrm d}}{\mathcalC^2}
  =M\Omega_0^2\sin^2\psi+ K\cos^2\psi,
\end{gather} 
where 
\begin{equation}
 \psi
 =\Omega_0 \tau-\arg C.
 \label{psi}
\end{equation}
Substituting here Eqs.~\eqref{S}, \eqref{B}, and simplifying results in
\begin{multline} 
  \frac{2\mathcal E_{\mathrm c}}{\mathcalC^2}
  +
  \frac{2\mathcal E_{\mathrm d}}{\mathcalC^2}
  =
  \left(
    \frac
    {\O_0^2\left(\TF\rho+M \sqrt{k\mathcal T-k\rho\v^2-\TF\rho\O_0^2}\right)}
    {\sqrt{k\TF-k\rho\v^2-\TF\rho\O_0^2}}
  \right.
  \\
  +
  \left(K-M\Omega_0^2+
2\sqrt{k\TF-k\rho\v^2-\TF\rho\O_0^2}\right)
\cos2\psi
  \Bigg).
\end{multline} 
The second term in the right-hand side of the last equation is zero according
to frequency equation 
\eqref{fr_eq-1-l}. Thus,
\begin{equation}
  \mathcal J=
    \frac
    {\mathcal C^2\O_0\left(\TF\rho+M \sqrt{k\mathcal T-k\rho\v^2-\TF\rho\O_0^2}\right)}
    {2\sqrt{k\TF-k\rho\v^2-\TF\rho\O_0^2}}.
\end{equation}
Now, postulating that 
the action $\mathcal J$ of the trapped wave  
is an adiabatic
invariant
\begin{equation}
 \mathcal J=\const
\end{equation}
clearly leads to Eqs.~\eqref{amp0-l-zzz}, \eqref{propto}:
\begin{equation}
  |\mathcal C|\propto
\sqrt{
  \frac{
    \sqrt{k\TF-k\rho\v^2-\TF\rho\O_0^2}
}{\O_0\left(\TF\rho+M \sqrt{k\mathcal T-k\rho\v^2-\TF\rho\O_0^2}\right)}}.
\label{J-moving}
\end{equation}

\section{Effective Hamiltonian system}
\label{sect-discuss}
In the previous section, we have shown that there is a
simple approach to solve a class of problems concerning strongly localized
oscillation in systems with time-varying parameters,
which 
is based on the calculation of the action of the trapped wave. Although we have observed
some difficulties arising from the ambiguity in selecting the appropriate energy
functional,
we have demonstrated that, in principle, it could be possible to calculate the evolution of the amplitude
for localized oscillation without direct asymptotic calculations.
However, 
based on empirical observations, we can propose an even simpler alternative for solving the same problem, which does not require calculating the action of the trapped wave.

Indeed, for both problems discussed in Sect.~\ref{sect-fixed} and \ref{sect-moving}, the
amplitude $\mathcal C(\epsilon t)$ is proportional to
$\sqrt{C_0(\epsilon t)}$; see Eqs.~\eqref{propto}, 
\eqref{C0C0}, whereas the latter quantity can be calculated without
considering the non-stationary perturbed problem. This is an 
unobvious result, which was obtained by the asymptotic analysis presented in
\cite{Gavrilov2024nody}.
%
The physical rationale for this observation becomes clear when recalling our
suggestion (Remark~\ref{remark-ham2}) to associate an effective single-degree-of-freedom Hamiltonian system with the discrete-continuous system under consideration.
%
%
Introduce the following equation:
\begin{gather}
\frac{\d }{\d t}
\left( \mathcal M(\et) \frac{\d \UF}{\d t} \right)
+ \mathcal K(\et) \UF = p(t).
\label{eff}
\end{gather}
For 
\begin{gather}
 \mathcal M(\et)=M(\et),
 \\
  \mathcal K(\epsilon t)= M(\epsilon t)\Omega_0^2(\epsilon t)
  \label{freq-harmonic-init}
\end{gather}
this equation describes the formal limiting case 
\begin{gather}
v=0,\qquad\mathcal T=0,\qquad \rho=0, \qquad k=0
\label{case-no-co}
\end{gather}
of an isolated mass-spring oscillator with time-varying
parameters defined by Eq.~\eqref{eq2} where
$P=0$. 
Accordingly, 
Eqs.~\eqref{CCCC},
\eqref{argC}--\eqref{phiOt}, 
\eqref{eq:s-ext-l0},
\eqref{amp0-l},
where Eq.~\eqref{case-no-co} is taken into account, express the solution of the
corresponding non-stationary unperturbed problem for Eq.~\eqref{eff}.
Analogously,  Eqs.~\eqref{sol-eg0}--\eqref{phi-def}, \eqref{propto}
\eqref{amp0-l-zzz} express the solution of the
corresponding non-stationary perturbed problem for Eq.~\eqref{eff}.
Consequently, both solutions satisfy
Eq.~\eqref{C0C0}. Now, we can substitute 
\begin{gather} 
  \mathcal M(\et)=\frac1{\Omega_0(\epsilon t) |C_0(\epsilon t)|}=
  \frac{\TF\rho+M \sqrt{k\mathcal T-k\rho\v^2-\TF\rho\O_0^2}}{
    \sqrt{k\TF-k\rho\v^2-\TF\rho\O_0^2}
}
  ,
  \label{M-eff}
  \\
  \mathcal K(\epsilon t)= \mathcal M(\epsilon t)\Omega_0^2(\epsilon t),
  \label{freq-harmonic}
\end{gather} 
where $|C_0(\et)|$ is defined by 
the right-hand side of Eq.~\eqref{amp0-l}
taken at $\et$. In the latter case, the above discussed formulae that express the solution of
the non-stationary problems for Eq.~\eqref{eff} transforms back into the
corresponding solutions of the problem formulated in Sect.~\ref{sect-form-moving}, 
inheriting the property~\eqref{C0C0} specific for the solutions of
Eq.~\eqref{eff}.

Equation \eqref{eff} can be reformulated in the Hamiltonian form
\begin{equation}
  \dot{\mathcal P}=-\pd{\mathcal H}{\mathcal Q}
  ,
  \qquad
  \dot{\mathcal Q}=\pd{\mathcal H}{\mathcal P},
  \label{ham-sys}
\end{equation}
where
\begin{equation}
  \mathcal H=\mathcal H(\mathcal Q,\mathcal P,t)\=\frac{\mathcal P^2}{2\mathcal M(\et)}+\frac {\mathcal K(\et)\mathcal Q^2}2
  -\mathcal Qp(t)
  \label{ham}
\end{equation}
is the corresponding Hamiltonian, 
\begin{equation}
\mathcal Q=\mathcal U,\qquad\mathcal P=\mathcal M\frac {\d\mathcal U}{\d t}
\label{QP}
\end{equation}
are the generalized
co-ordinate and the generalized impulse, respectively.
We call Eqs.~\eqref{ham-sys}--\eqref{QP} where the effective mass $\mathcal M$ and effective stiffness 
$\mathcal K$ are expressed by Eqs.~\eqref{M-eff}--\eqref{freq-harmonic} the
effective Hamiltonian system for the trapped mode discussed in Sect.~\ref{sect-moving}.
Note that mass $\mathcal M$ is a positive quantity 
\begin{equation}
\mathcal M>0
\label{mM>0}
\end{equation}
due to Eqs.~\eqref{restr-par-21}, \eqref{restr-par-22}, \eqref{fr_eq-1-l}.
The frequency $\Omega_0$ satisfies both the frequency equation 
\eqref{fr_eq-1-l} and that of the harmonic oscillator \eqref{freq-harmonic}. 
The effective parameters
$\mathcal M$ and $\mathcal K$ can be calculated without considering the non-stationary perturbed
problem since they depend on $|C_0(\et)|$ and $\Omega_0(\et)$ only.
\begin{remark} 
Equation \eqref{eff} is investigated in detail in Appendix~\ref{App-1d}.
\end{remark} 
\begin{remark} 
Due to the frequency equation 
\eqref{fr_eq-1-l}, Eq.~\eqref{M-eff}
can be
equivalently rewritten as follows:
\begin{equation}
\label{amp01}
\mathcal M=
    \frac{M^2\O_0^2-K M +2\TF \rho}{M\O_0^2 - K}
.
\end{equation}
\end{remark}


\begin{remark} 
Newton's second law in the form of Eq.~\eqref{eff} governs the dynamics of
a variable-mass particle, i.e., an open system with mass supply. 
The current formulation assumes zero momentum supply; i.e., the external mass impacts the system with zero velocity in the current reference frame
\cite{Irschik2004,LeviCivita1928,LeviCivita1928a}. If, instead, we assume that the relative velocity between impacting mass and the variable-mass point under consideration is zero, then the corresponding Newton's law is expressed by the different equation:
\begin{gather}
\mathcal M(\et) \frac{\d^2 \UF}{\d t^2}
+ \mathcal K(\et) \UF = p(t).
\label{eff2}
\end{gather}
Equations \eqref{eff} and \eqref{eff2} describe the dynamics of two distinct open systems; therefore, they do not admit a standard Lagrangian formulation.
However, both of them admit a Hamiltonian formulation with the corresponding
Hamiltonians given by \eqref{ham} and 
\begin{equation}
\mathcal H=\hat{\mathcal H}(\mathcal Q,\mathcal P,t)\=\frac{\mathcal P^2}{2}+\frac {\mathcal K(\et)\mathcal Q^2}{2\mathcal M(\et)}
  -\frac{\mathcal Qp(t)}{\mathcal M(\et)},
  \label{ham2}
\end{equation}
respectively. 
In Eq.~\eqref{ham2},
\begin{equation}
  \mathcal Q={\mathcal U},\qquad\mathcal P=\frac {\d\mathcal U}{\d t}.
\end{equation}
For $t>t_0$, the corresponding actions in the zeroth-order approximation are $\mathcal H/\Omega_0$,
which equal to
\begin{equation}
  \mathcal J=\frac{\mathcal M \Omega_0 \mathcal C^2}2
  \label{J2nd}
\end{equation}
and
\begin{equation}
  \hat{\mathcal J}=\frac{ \Omega_0 \mathcal C^2}2,
\end{equation}
respectively. Here, $\mathcal C$ is the oscillation amplitude. Assuming the
adiabatic invariance of the corresponding actions leads to 
\begin{equation}
  \mathcal C\propto \frac1{\sqrt{\mathcal M \Omega_0}}=\frac{1}{\sqrt[4]{\mathcal K\mathcal M}}
\label{good-prop}
\end{equation}
and
\begin{equation}
\mathcal C\propto \frac1{\sqrt{\Omega_0}}=\sqrt[4]{\frac{\mathcal M}{\mathcal K}},
\label{bad-prop}
\end{equation}
respectively. The classical treatment, see, e.g., \cite{Arnold2009}, corresponds to the case when 
$\mathcal M=\const$, and, therefore, the last two equations are proportional to each other with
a constant coefficient of proportionality.
\end{remark} 
\begin{remark} 
  Note that Eq.~\eqref{good-prop} coincides with Eq.~\eqref{J-moving} if notation
  \eqref{M-eff} is accepted. 
  Hence, the same adiabatic invariant governs the evolution of both the trapped mode amplitude and  
  the amplitude of the corresponding solution for the effective Hamiltonian system.
\end{remark} 

Assuming that Eq.~\eqref{C0C0} holds for problems more complicated than those considered in this paper, one may attempt to derive the corresponding solutions; see the Conclusion for further discussion.



\section{Conclusion}
In the paper, we have addressed the problem of the non-stationary dynamics of the 
discrete-continuous system with time-varying
parameters, which was previously investigated using asymptotic methods
in \cite{Gavrilov2024nody}.
We have demonstrated the adiabatic invariance of the quantity
that we have called the action of the trapped wave; see Sect.~\ref{sect-adibatic-fixed}.
The action is the ratio
of the conserved finite total energy of the trapped wave to its frequency.
To show the adiabatic invariance, we have used the formal asymptotic solution
obtained in \cite{Gavrilov2024nody} by the space-time ray method. Finally, we
have established an analogy between the dynamics of the discrete-continuous system where a single trapped wave can
exist and  single-degree-of-freedom Hamiltonian systems. For the latter, the adiabatic invariance
of the action is a well-known fact.

Derivation of the formal asymptotic solution for discrete-continuous 
systems is a non-trivial task. Apart from the extensive
calculations, a challenge lies 
in the {\it a priori} construction of a suitable asymptotic ansatz 
that captures the dynamics of the perturbed system.
Postulating the adiabatic invariance of the
action of the trapped wave is a less laborious approach.
However, this approach also requires an {\it a priori} selection of the energy functional used to calculate the action.
This challenge was addressed in Sect.~\ref{sect-adibatic-moving}, in which we assume that the discrete inclusion moves along the continuous sub-system.

Finally, in Sect.~\ref{sect-discuss}, we have presented a highly
straightforward approach to deriving the amplitude evolution law, which
relates it to the square root of the unperturbed system's response to the
pulse excitation $\delta(t)$ formally evaluated at $\epsilon t$.
This remarkable property \eqref{C0C0} is inherited from the Hamiltonian system
\eqref{ham-sys}, \eqref{ham}, which possesses the same adiabatic invariant
(the action) as the trapped wave.
Postulating this
property allows us to solve the problem without any {\it a priori} construction of the ansatz or the energy functional.
While a formal justification for extending the result in Eq.~\eqref{C0C0} to systems beyond those considered in \cite{Gavrilov2024nody} and the present study is not yet established, our preliminary findings suggest that such a generalization is plausible.
In a subsequent publication, we demonstrate that this observation remains valid for an analogous problem where the taut string is replaced by an Euler-Bernoulli beam, at least for the case of a fixed inclusion.
A very particular case of such a problem,
with only one time-varying parameter, is discussed in our previous study \cite{Shishkina2019jsv}. 
Conversely, the relationship \eqref{C0C0} does not generally hold for Hamiltonian systems with multiple time-varying parameters.
For instance, Eq.~\eqref{C0C0} is not fulfilled for
the aforementioned system
\eqref{ham-sys}, \eqref{ham2} due to Eqs.~\eqref{good-prop}, \eqref{bad-prop},
since the unperturbed problems for systems described by Eqs.~\eqref{ham-sys},
\eqref{ham} and Eqs.~\eqref{ham-sys}, \eqref{ham2} coincide. 

While the proposed approaches facilitate the analysis of problems similar to the one discussed in this paper, we are not currently aware of any validation method distinct from the original asymptotic framework.
However, to proceed with such a validation, we again need 
to {\it a priori} construct the asymptotic ansatz and the energy functional necessary to
calculate the action. The similar position was presented in
\cite{Babich2002} concerning the validation of the Whitham theory.

It is also important to highlight the differences between the problems considered in
this paper and those addressed within the Whitham theory framework 
\cite{Whitham1999,Buehler2009,Bretherton1966,Whitham1965,Whitham1967,Lighthill1967full}.
Unlike the Whitham theory, which assumes a wave-train structure for the
solution at all stages of the motion, our work concerns an initial value
problem and a pulse excitation. 
Furthermore, our approach treats energy as a conserved, finite quantity, whereas the Whitham theory relies on an averaged energy density, as the total energy of a wave-train is inherently divergent.
Thus, the action of a trapped wave is conserved over time, whereas within the framework of Whitham theory, we can only consider the corresponding density, which satisfies a conservation law in the form of a PDE.

The proposed simplified approaches can be used in engineering applications.
We anticipate that they will be particularly effective for investigating the vibration of a beam on an elastic foundation subjected to a non-uniformly moving inertial load.

%

\section*{Acknowledgement}
The authors are grateful to A.M.~Krivtsov, Yu.A.~Mochalova for discussions. 

\section*{Funding}
The research is supported by the Russian Science Foundation
(project 26-11-00372).

\appendix

\section{Calculation of the false invariant $J^{(1)}$}
\label{App-BAD}
Here, we calculate the false invariant $J^{(1)}$ defined by Eq.~\eqref{Inv-1}.
Using Eqs.~\eqref{eq:s-ext-l}, \eqref{energy-unpert-c}, \eqref{energy-unpert-d1},
\eqref{energy-unpert-new}--%
\eqref{conf-force}, one obtains:
\begin{multline}
  \frac{2\tilde E^{(1)}\{\bar u\}}{\mathcalC^2}=\left(K -\Omega_0^2\left(M+\frac{2\TF \rho}{\sqrt{k\TF - k\rho v^2 -\TF \rho \Omega^2}}\right)\right)\cos2\psi
		+\frac{4k(\TF-\rho v^2)}{\sqrt{k\TF - k\rho v^2 -\TF \rho \Omega^2}}\cos^2\psi
	\\
	+
	\frac{K\TF-\rho v^2(K+M\Omega_0^2)}{\TF-\rho v^2}
		+\frac{\TF \Omega_0^2}{\TF-\rho v^2}\left( M + \frac{2\rho^2 v^2}{\sqrt{k\TF - k\rho v^2 -\TF \rho \Omega^2}}\right)
	.
	\label{full-energy-v}
\end{multline}
Here, $\psi$ is defined by Eq.~\eqref{psi}.
Using  frequency equation~\eqref{fr_eq-1-l}, one can demonstrate that
\begin{equation}
	K -\Omega_0^2\left(M+\frac{2\TF \rho}{\sqrt{k\TF - k\rho v^2 -\TF \rho \Omega^2}}\right)=-\frac{2k(\TF-\rho v^2)}{\sqrt{k\TF - k\rho v^2 -\TF \rho \Omega^2}}.
\end{equation}
Hence, Eq.~\eqref{full-energy-v} transforms as follows:
\begin{multline}
  \frac{2\tilde E^{(1)}}{\mathcalC^2}=\frac{2k(\TF-\rho v^2)}{\sqrt{k\TF - k\rho v^2 -\TF \rho \Omega^2}}
		+
	\frac{K\TF-\rho v^2(K+M\Omega_0^2)}{\TF-\rho v^2}
		+\frac{\TF \Omega_0^2}{\TF-\rho v^2}\left( M + \frac{2\rho^2 v^2}{\sqrt{k\TF - k\rho v^2 -\TF \rho \Omega^2}}\right)
	\\
	=
	K+M\Omega_0^2 + \frac{2k(\TF-\rho v^2)+2\rho^2 v^2 \TF \Omega_0^2}{(\TF -\rho v^2)\sqrt{k\TF - k\rho v^2 -\TF \rho \Omega^2}}
  .
  \label{Utmp}
\end{multline}
The following algebraic identity is true:
\begin{gather}
	2k(\TF-\rho v^2)+2\rho^2 v^2 \TF \Omega_0^2
	=2(\TF -\rho v^2)(k\TF - k\rho v^2 -\TF \rho \Omega^2)
	+ 2\TF^2\rho \Omega_0^2.
  \label{identi}
\end{gather}
Thus, substituting Eq.~\eqref{identi} into the second term in the right-hand
side of Eq.~\eqref{Utmp} and 
using frequency equation~\eqref{fr_eq-1-l} results in
\begin{equation}
	\tilde E^{(1)}={\mathcalC}^2\Bigg(M\Omega_0^2+\frac{\TF^2\rho\Omega_0^2}{(\TF-\rho v^2)\sqrt{k\TF - k\rho v^2 -\TF \rho \Omega^2}}\Bigg).
	\label{full-en-v}
\end{equation}
Substituting this expression into Eq.~\eqref{Inv-1}, one gets:
\begin{equation}
	J^{(1)}={\mathcal C}^2\,\frac{\Omega_0(M(\TF-\rho v^2)\sqrt{k\TF - k\rho v^2 -\TF \rho \Omega^2}+\TF^2\rho)}{(\TF-\rho v^2)\sqrt{k\TF - k\rho v^2 -\TF \rho \Omega^2}}.
\end{equation}
Thus, instead of Eq.~\eqref{J-moving}, one obtains the following expression for
the oscillation amplitude:
\begin{equation}
	|\mathcal C|\propto
	\sqrt{\frac{(\TF -\rho v^2)\sqrt{k\TF - k\rho v^2 - \TF \rho \Omega^2}}{\Omega_0(M(\TF-\rho v^2)\sqrt{k\TF - k\rho v^2 -\TF \rho \Omega^2}+\TF^2\rho)}},
		\label{propto-2}
\end{equation}
which is in a contradiction with asymptotically correct formula 
\eqref{amp0-l-zzz}.
\begin{remark} 
  It is easy to see that postulating $J^{(2)}$ to be an adiabatic invariant
  also does not allow us to obtain  formula 
\eqref{amp0-l-zzz}.
\end{remark} 

\section{Oscillation of a mass-spring system with time-varying parameters}
\label{App-1d}
Here, we consider Eq.~\eqref{eff}.
\subsection{Stationary unperturbed problem}
Firstly, consider the corresponding stationary ($p=0$) unperturbed ($\epsilon=0$) problem. We have
\begin{equation}
\mathcal K=\const,
\qquad
\mathcal M=\const.
\label{all-constants-eff}
\end{equation}
The solution is
\begin{equation}
  \mathcal U=\bar{\mathcal U}\=|C|
\cos(\Omega_0 t-\arg C),
\label{eq:solution_inhomogeneous_stationary_problem-ext-eff}
\end{equation}
where $C$ is an arbitrary complex constant,
\begin{equation}
  \Omega_0=\sqrt{\frac{\mathcal K}{\mathcal M}}.
  \label{freq-eq-eff}
\end{equation}

\subsection{Non-stationary unperturbed problem}
Next, consider the non-stationary problem with $p\neq0$ in the zeroth-order approximation ($\epsilon=0$), i.e., the problem parameters 
\begin{equation}
\mathcal K=\mathcal K(0),
\qquad
\mathcal M=\mathcal M(0)
\label{all-slow0-eff}
\end{equation}
are constants,
which are equal to the corresponding initial values at $t=0$.

Put $\epsilon=0$ and consider the following equation:
\begin{gather}
 \mathcal M(0) \frac{\d^2 \UF}{\d t^2}
+ 2\gamma \dd{\UF}{t}
+ \mathcal K(0) \UF = p(t).
\label{eff0-gamma}
\end{gather}
In accordance with the limit absorption principle, we introduce a dissipative term 
$2\gamma \dd{\UF}{t}$ 
into the left-hand side of Eq.~\eqref{eff0-gamma} to enable the application of the classical Fourier transform.

Applying the Fourier transform yields:
\begin{gather}
 -\mathcal M(0)\Omega^2 \UF_F
- 2\I\Omega\gamma \UF_F
+ \mathcal K(0) \UF_F = p_F(\Omega),
\\
\UF_F=\int_{-\infty}^{+\infty}\UF\EXP{\I\Omega t}\,\d t.
\end{gather}
Thus,
\begin{equation}
  \UF=\frac1{2\pi}\int_0^\infty\frac{p_F(\Omega)\EXP{-\I\Omega t}\,\d\Omega}{-\mathcal M(0)\Omega^2-2\I\gamma\Omega+\mathcal K(0)}+\cc,
\end{equation}
where $\cc$ are the corresponding complex conjugate terms.
In the conservative case $\gamma=+0$, and we have:
\begin{multline}
  \UF=
  \frac1{2\pi}\lim_{\gamma\to+0}
  \int_0^\infty\frac{p_F(\Omega)\EXP{-\I\Omega t}\,\d\Omega}{-\mathcal M(0)\Omega^2-2\I\gamma\Omega+\mathcal K(0)}+\cc
  =
  \frac{-1}{2\pi \mathcal M(0)}\lim_{\gamma\to+0}
  \int_0^\infty\frac{p_F(\Omega)\EXP{-\I\Omega t}\,\d\Omega}{\Omega^2+2\I\gamma\Omega-\Omega_0^2(0)}+\cc
  \\= 
  \frac{1}{4\pi \mathcal M(0) \Omega_0(0)}
  \int_{0}^\infty 
  \frac{p_F(\Omega)\EXP{-\I\Omega t}}{\Omega+\I0-\Omega_0(0)}
  \,\d \Omega+\cc+O(t^{-\infty})
  =
  \frac{\I p_F\big(\Omega_0(0)\big)\EXP{-\I\Omega_0(0)t}}{2\mathcal
  M(0)\Omega_0(0)}+\cc+O(t^{-\infty}), \quad t\to\infty.
  \label{int-calc}
\end{multline}
Here, Eq.~\eqref{mM>0} has been taken into account.
The integral in 
 Eq.~\eqref{int-calc}
is estimated by the stationary phase method \cite{Fedoryuk1977}, using
Sokhotski-Plemelj theorem for the real line or by transforming the contour to the closed one by the
Jordan lemma. Finally, this yields:
\begin{equation}
  \UF=|C(0)|\cos\big(\I\phi-\arg C(0)\big)+O(t^{-\infty}),
  \quad 
  t\to\infty.
\end{equation}
Here,
\begin{gather}
|C(0)|=\big|C_0(0)p_F\big(\O_0(0)\big)\big|,
\label{CCCC-eff}
\\
|C_0(0)|=
\left.
\frac1{\mathcal M\Omega_0}
  \right|_{t=0},
\label{amp0-eff}
\\
\arg C(0)=\arg p_F\big({\O_{0}(0)}\big)
+
\arg C_0
,
\label{argC-eff}
\\
\arg C_0(0)=
\frac\pi2
,
\label{Co-value-eff}
\\
\phi=-\I \Omega_0(0) t.
\label{phiOt-eff}
\end{gather}
Since $p$ satisfies 
Eq.~\eqref{pulse}, and the function $\cos\big(\Omega_0(0)t\big)$ satisfies the homogeneous
equation that corresponds to Eq.~\eqref{eff0-gamma}, we have
\begin{equation}
  \UF=|C(0)|\cos\big(\I \phi-\arg C(0)\big),
  \quad 
  t>t_0.
\label{UFF-eff}
\end{equation}

\subsection{WKB analysis of the non-stationary perturbed problem}

Based on the formal framework provided in \cite{nayfeh2008perturbation}, 
for times
\eqref{tt0>},
we investigate the free non-stationary oscillation of the mass-spring system
defined by Eq.~\eqref{eff}.

We introduce {the} slow time $T=\epsilon t$ and 
represent the solution in the form
of the following ansatz 
%
\begin{gather}
\mathcal U=\mathcal W\,\exp\phi+\cc,
\label{App1-U-phi}
\end{gather}
where 
\begin{gather}
\mathcal W(T)=\sum_i \epsilon^i\mathcal W_i(T),
\label{W-eff}
\\
\phi(T)=-\I\int_0^t\Omega_0(\epsilon \hat t)\,\d\hat t=
-\I\int_0^t\sqrt{\tfrac{\mathcal K(\epsilon \hat t)}{\mathcal M(\epsilon \hat t)}}\,\d\hat t
\label{phip-eff}
\end{gather}
are the amplitude and the phase, respectively.
Then, we consider 
$T$
and $\phi$ as independent time-like variables, and use the corresponding representations 
\begin{equation}
\label{dot-xt}
\begin{aligned}
&\frac{\d}{\d t}=-\I \Omega_0 \PARTIAL_\phi' + \epsilon \PARTIAL_T'
+O(\epsilon),\\
&\frac{\d^2}{\d t^2}=-\Omega_0^2 \PARTIAL_{\phi\phi}'' - 2\epsilon \I \Omega_0 \PARTIAL_{\phi T}''
-\epsilon \I \Omega_0{}'_T \PARTIAL_\phi' +O(\epsilon^2)
\end{aligned}
\end{equation}
for differential operators with respect to time.
Substituting Eqs.~\eqref{App1-U-phi}--\eqref{dot-xt} into 
Eq.~\eqref{eff} results in the following first approximation equation for 
the leading-order term $\mathcal W_0(T)$:
\begin{equation}
2\sqrt{\mathcal M \mathcal K}\,{\mathcal W}_0{}'_T+\mathcal M\left(\sqrt{\frac{\mathcal K(T)}{\mathcal M(T)}}\,\right)'_T\mathcal W_0
+{\mathcal M}{}'_T\sqrt{\frac{\mathcal K(T)}{\mathcal M(T)}} \mathcal W_0
=0.
\label{long-P}
\end{equation}
Taking into account that 
\begin{equation}
\left(\sqrt{\frac{\mathcal K(T)}{\mathcal M(T)}}\,\right)'_T
=
\frac{\mathcal K'_T}{2\sqrt{\mathcal K \mathcal M}}-\frac{\mathcal M'_T\sqrt \mathcal K}{2\mathcal M^{3/2}},
\end{equation}
after simplification, one can rewrite 
Eq.~\eqref{long-P} in the following form:
\begin{equation}
\frac{\mathcal W_0{}'_T}{\mathcal W_0}=
-\frac{\mathcal K'_T}{4\mathcal K}-\frac{\mathcal M'_T}{4\mathcal M}
.
\label{App1-1st-P}
\end{equation}
The solution is
\begin{equation}
    \WF_{0}(T)
  =\frac {\breve C}
  {\sqrt[4]{\mathcal M(T) \mathcal K(T)}}
  = \frac 
  {\breve C}
  {\sqrt{\mathcal M(T)\Omega_0(T)}},
\label{App1-sol-0}
\end{equation}
where $\breve C$ is an arbitrary complex constant. Then we require that for
$T=0$, expressions for $\mathcal U$ 
\eqref{UFF-eff}
and 
\eqref{App1-U-phi} are identical. This is the matching condition that
yields 
\begin{equation}
  \breve C=\frac{C(0)}2.
\end{equation}
Thus,
\begin{gather}
\UF(t) =
|\mathcal C(\epsilon t)|
\cos\big(\I\phi - \arg \mathcal C\big)+\text{o}(1), 
\label{sol-eg0-eff}
\\
|\mathcal C|=|\mathcal C(\epsilon t)|=|C(0)|{\frac{|\mathcal C_0(\epsilon t)|}{|\mathcal C_0(0)|}},
\label{amp-eff}
\\
\arg \mathcal C=\arg C(0),\qquad
\arg \mathcal C_0=\arg C_0(0),
\\
\phi=
{-}\I\int_{0}^{t}
 \Omega_0\big(\epsilon\hat t\big)\,\d \hat t,
\label{phi-def-eff}
\\
|\mathcal C_0|=
|\mathcal C_0(\epsilon t)|\=
\frac1{\sqrt{\mathcal M\Omega_0}}.
\label{amp-expl-eff}
\end{gather}
One can see that Eqs.~\eqref{sol-eg0-eff}--\eqref{amp-expl-eff} have the same
structure as Eqs.~\eqref{sol-eg0}--\eqref{amp-expl} and relationships
\begin{equation}
  |\mathcal C(\et)|\propto|\mathcal C_0(\et)|
    ,\qquad
    |\mathcal C_0(\epsilon t)|=\sqrt{|C_0(\epsilon t)|}
  \label{propto-eff}
\end{equation}
are again fulfilled.

In the particular case $\dot {\mathcal M}=0$, the amplitude $\mathcal W_0$
is proportional to the inverse of the square root of the natural frequency.
\begin{equation}
\mathcal W_0(T)=\frac {\bar {\mathcal C}}{\sqrt{\Omega_0(T)}},
\label{L-G}
\end{equation}
where $\bar{\mathcal C}=\mathcal C/\sqrt{\mathcal M}$ is an arbitrary constant.
This is the classical Liouville--Green (or WKB) approximation \cite{nayfeh2008perturbation}.

\subsection{Adiabatic invariance of the action}
According to Eqs.~\eqref{amp-eff}, \eqref{amp-expl-eff},
\begin{equation}
 I=
 \frac{|\mathcal C|}{{|\mathcal C_0|}}
 =
|\mathcal C|
\sqrt{\mathcal M\Omega_0
}
 =\frac{|C(0)|}{|\mathcal C_0(0)|}
=
\const,
\label{I-intro-eff}
\end{equation} 
for times
\eqref{tt0>}.
Thus, according to the asymptotic calculations above,
we have found that $I$ or any function $f(I)$ of the single variable $I$
such that $f'\neq0$ 
is an adiabatic invariant.

On the other hand, 
within the framework of Hamiltonian mechanics \cite{Arnold2009}, it is established that for a single-degree-of-freedom system with Hamiltonian \eqref{ham} and $t>t_0$, the action variable $\mathcal J$ of the action--angle pair is an adiabatic invariant:
\begin{gather}
 \mathcal J=
 \left. 
   \frac{\mathcal H\{\bar {\mathcal U}\}}{\Omega_0}
 \right|_{C=\mathcal C},
 \\
 \mathcal H\{\bar {\mathcal U}\}=
 \left.
 \frac{\mathcal P^2}{2\mathcal M(\et)}+\frac {\mathcal K(\et)\mathcal Q^2}2
 \right|_{\mathcal Q=\bar {\mathcal U},\ \mathcal P=
 \mathcal M\frac {\d \bar{\mathcal U}}{\d t}},
  \label{ham-eff}
\end{gather}
where $\bar{\mathcal U}$ is defined by 
Eq.~\eqref{eq:solution_inhomogeneous_stationary_problem-ext-eff}.
We have 
\begin{equation}
  \mathcal H\{\bar{\mathcal U}\}=\frac{C^2\mathcal M^2\Omega_0^2\sin^2\psi }{2\mathcal M}+\frac{C^2\mathcal K^2\cos^2\psi}2=\frac{C^2\mathcal M\Omega_0^2}{2},
\end{equation}
where Eq.~\eqref{freq-eq-eff} is taken into account, and
$\psi=\Omega_0t-\arg C$.
Thus, the action variable is
\begin{equation}
\mathcal J=\frac{\mathcal C^2\mathcal M\Omega_0}2=\frac {I^2}2.
\end{equation}

It can be seen that assuming $\mathcal J=\const$ leads to Eq.~\eqref{propto-eff}, which was previously obtained through direct asymptotic calculations.

\bibliographystyle{plainnat}
\bibliography{bib/serge-gost,bib/all}

\end{document}